\documentclass[%
 reprint,
 amsmath,amssymb,
 aps,
]{revtex4-2}

\usepackage{graphicx}
\usepackage{dcolumn}
\usepackage{bm}
\usepackage{braket}
\usepackage{nicematrix}
\usepackage{hyperref}
\usepackage{tikz}
\usepackage{cleveref}
\begin{document}

\preprint{APS/123-QED}

\title{Superconductivity Near a Quantum Critical Point: Bounds on the Transition Temperature in the $\gamma$-Model}

\author{Ahmed Elezaby}
\author{Artem Abanov}
\affiliation{Department of Physics and Astronomy, Texas A\&M University, College Station, TX 77843, USA}

\date{\today}

\begin{abstract}
Near a quantum critical point (QCP) in a metal, strong Fermion-Fermion interactions mediated by soft collective bosons give rise to two competing phenomena: non-Fermi liquid behavior and superconductivity that deviates from conventional BCS and Migdal-Eliashberg theories. We consider the problem of obtaining closed-form analytical lower and upper bounds on transition temperatures for such systems. We focus mainly on a class of models known as the $\gamma$-model, a variation of the Eliashberg theory of Superconductivity where the effective interaction potential scales as $V(\Omega) \propto 1/|\Omega|^\gamma$. Building on a recent reformulation of Migdal-Eliashberg theory—expressed as a classical infinite spin chain with nonlocal interactions \cite{emil1,emil2}—and employing a linear algebra analysis of the Hessian matrix obtained from the Free Energy functional, we derive rigorous, closed-form expressions for upper and lower bounds on the superconducting transition temperature for any $\gamma > 0$. The main result of the paper is to establish an analytical upper bound on the transition temperature in closed form. 
Our upper bound is significantly tighter than those currently available in the literature and demonstrates rapid convergence toward results from prior numerical studies. Also, by applying the singularity condition directly to the unbounded Hessian matrix,
our independently performed calculations confirm the lower bounds previously established in the
literature \cite{Kiessling2025}.
\end{abstract}

\maketitle

\section{\label{sec:intro} Introduction}

Superconductivity is a macroscopic quantum state of matter in which certain materials exhibit precisely zero electrical resistance when cooled below a characteristic critical temperature, $T_c$. The field's origin traces back to 1911, when Dutch physicist H. Kamerlingh Onnes, in his quest to understand the behavior of matter near absolute zero, first observed superconductivity in mercury \cite{Onnes1911}. This landmark discovery was enabled by his earlier success in liquefying helium in 1908 \cite{Onnes1908}, which provided the cryogenic environment necessary to reach 4.2 K. At this temperature, Onnes found that the electrical resistance of his mercury sample vanished completely. Another defining feature is perfect diamagnetism below $T_c$, known as Meissner effect, and was discovered in 1933 by  Walther Meissner and Robert Ochsenfeld \cite{Meissner1933}. This demonstrated that superconductivity is a true thermodynamic equilibrium state, not merely a dynamic state of infinite
conductivity, thereby revealing the deep quantum mechanical nature of the phenomenon.

The first successful theoretical microscopic description came more than 20 years later by Bardeen, Cooper, and Schrieffer (BCS) \cite{BCS,Cooper1}. BCS theory's key idea is that a phonon-mediated effective attraction overcomes the Coulomb repulsion, binding electrons into Cooper pairs \cite{Cooper1,BCS}. These pairs exhibit unique properties. Rather than behaving as independent particles, they act as a collective entity exhibiting a macroscopic wavefunction that extends across the entire material. This coherent quantum state is characterized by an energy gap ($\Delta$), representing the binding energy of the pairs, which prevents low-energy excitations and underpins the system’s remarkable properties. BCS theory was a monumental success, explaining the energy gap, the $T_c$ formula, the specific heat jump, and the isotope effect \cite{isotope1,isotope2}.

Despite its success, BCS is a weak-coupling theory. It was generalized by Eliashberg (building on work by Migdal \cite{Migdal1958}) in 1960 to handle the strong-coupling regime, where the electron-phonon interaction is large \cite{eliash1,eliash2}. Eliashberg theory explicitly accounts for the retardation of the phonon-mediated interaction—the fact that the lattice response (phonons) is much slower than the electronic motion. This strong-coupling theory, which requires solving a set of coupled integral equations, provided a quantitative description for materials like lead and mercury, where BCS theory failed numerically. 

This phonon-mediated paradigm reigned until 1986, with the discovery of high-$T_c$ superconductivity in cuprates by Bednorz and Müller \cite{Bednorz1986}. The transition temperatures observed (eventually exceeding 130 K) were far beyond what was believed possible within the phonon-mediated Eliashberg framework \cite{Schilling1993,Scalapino1995}. This discovery ushered in the era of unconventional superconductivity. These materials, and subsequently others like iron-based pnictides \cite{Kamihara2008} and heavy-fermion compounds \cite{Stewart1984}, share a common trait: superconductivity often emerges on the border of another ordered phase, typically antiferromagnetism.
This proximity to a magnetic phase suggests a different pairing mechanism. Instead of phonons, the glue binding the Cooper pairs is thought to be the collective fluctuations of the competing order. When this competing order (e.g., magnetism) is tuned to zero temperature by a non-thermal parameter (like pressure, doping, or magnetic field), a Quantum Critical Point (QCP) is realized \cite{Hertz1976}. At a QCP, the system is dominated by scale-invariant quantum fluctuations. These gapless, critical fluctuations (e.g., spin fluctuations) provide a powerful, 
universal
pairing interaction \cite{Monthoux1992,Scalapino2012}.

This new paradigm presents profound theoretical challenges. The pairing interaction is no longer gapped. There is no large energy scale separation that allowed for the Cooper logarithm to be the major contribution. The bosonic mode cannot be regarded as slow, 
causing a breakdown of Migdal's theorem, which is fundamental to Eliashberg theory. The system's normal state is often a "strange metal" or non-Fermi liquid, lacking the well-defined quasiparticles that form the basis of BCS \cite{Chubukov2020}. The so-called "gamma model" (or $\gamma$-model) has emerged as a key theoretical framework to address this regime \cite{abanovnormal,MoonChubukov2010,paper1,paper2,paper3,paper4,paper5,paper6}. It simplifies the problem by modelling the pairing interaction mediated by these critical fluctuations with a phenomenological, singular frequency dependence: $V(\Omega) \propto 1/|\Omega|^\gamma$. The exponent $\gamma$ parameterizes the nature of the critical fluctuations. This model allows for a focused study of how a singular interaction gives rise to superconductivity and determines $T_c$, bridging the gap between quantum criticality and pairing. For a list of quantum-critical materials that are effectively described by the \( \gamma \)-model, we refer the reader to \cite{paper1}.

Alongside these developments in quantum critical pairing, the calculation of bounds on the transition temperature, $T_c$, has emerged as a general area of investigation within electron-boson models. Recent studies have primarily focused on establishing these limits within standard Eliashberg and electron-phonon frameworks. For instance, \cite{Esterlis2018} derived bounds on $T_c$ by constraining the pairing vertex, while subsequent works have evaluated upper limits in the strong-coupling regime \cite{Gnezdilov2025} and provided analytical constraints within Eliashberg–McMillan theory \cite{Sadovskii2024, Sadovskii2025}. Here, we focus exclusively on evaluating the upper and lower bounds of $T_c$ within the specific mathematical structure of the $\gamma$ model, determining how its unique, generalized scaling features dictate the limits of the transition temperature.

\subsection{Universality of the $\gamma$-Model}

The $\gamma$-model is not just an abstract mathematical construct; it also provides a mathematical bridge between different physical systems, with the exponent $\gamma$ acting as a tuning parameter that can recover the essential mathematical structures and physical properties of both BCS and Eliashberg theories. The BCS model assumes a frequency-independent (non-retarded) interaction, $V(\Omega) = V_0$, which is active up to a cutoff $\omega_D$. This is mathematically equivalent to setting $\gamma = 0$ in the gamma model, $V(\Omega) \propto 1/|\Omega|^0 = \text{constant}$. In this limit, the gamma model's gap equation (with an appropriate cutoff) reduces to the BCS gap equation. The $T_c$ calculation then yields the famous non-perturbative BCS formula, $T_c \propto \omega_D \exp(-1/\lambda)$, where $\lambda$ is the dimensionless electron-phonon coupling constant.

The standard Eliashberg theory for phonon-mediated superconductivity is a fully retarded theory. The electron-phonon kernel $\lambda(\Omega)$ (which describes the interaction) decays rapidly at high frequencies. Specifically, for $\Omega \gg \omega_D$ (the maximum phonon frequency), the kernel has the asymptotic behavior $\lambda(\Omega) \propto 1/\Omega^2$. This is precisely the case of the gamma model with $\gamma = 2$. This value represents a safe case where the Matsubara sums converge quickly, just as they do in Eliashberg theory. Thus, the $\gamma = 2$ model captures the essential mathematical character (the convergence and scaling) of the conventional, retarded phonon-pairing problem.
The gamma model, therefore, interpolates between the non-retarded BCS-like problem ($\gamma = 0$) and the retarded Eliashberg-like problem ($\gamma = 2$). It also generalizes to different retarded and non-retarded regimes for different values of $\gamma$, providing a universal framework that describes general quantum critical behavior, not just quantum critical superconductivity. 

For any $\gamma>0$, the gamma model requires no upper frequency cutoff. As such, the model has only one energy scale --- the bare electron-boson coupling constant $g$. As a consequence, every other energy scale, such as $\Delta$ or $T_c$ must have a form $gf(\gamma)$, where $f(\gamma)$ is a universal function of the dimensionless parameter $\gamma$. This simple observation makes the gamma model a universal model of a metal's behaviour at QCP. 

A recent framework for analyzing the strong-coupling Migdal-Eliashberg theory, and its $\gamma$-model generalization, was developed in the work of Ref.~\cite{emil1,emil2}, where the theory was mapped onto a classical spin chain with long-range Heisenberg exchange and Zeeman terms. In this representation, the discrete fermionic Matsubara frequencies, $\omega_m = (2m+1)\pi T$, become the sites of a one-dimensional chain. A three-component classical spin vector, $\boldsymbol{S}_m$, is placed at each site, with its orientation encoding the properties of the superconducting order parameter at that frequency. This reduces the problem of solving complex integral equations to the optimization of a well-defined free energy functional, expressed entirely in terms of these spin variables. Within this framework, the analysis becomes significantly more tractable; in particular, the critical transition temperatures, which mark the instability of the normal state, are identified by performing a linear stability analysis. This translates directly to calculating the eigenvalues of the Hessian matrix (the second functional derivative) of the free energy. A transition occurs when an eigenvalue crosses zero, signaling the emergence of a new state. As we shall show in Sec. \ref{sec:gammamodel}, the linearzied Hessian matrix $H$ at the normal state
 reads: 

       \begin{equation}\label{eq:intro:Hessiandiag}
     H_{nn} = (2n+1)\tau^\gamma + 2 \sum_{m=1}^{n} \frac{1}{m^\gamma} - \frac{1}{(2n+1)^\gamma},
\end{equation}
and for \( m \neq n \):
\begin{equation}\label{eq:intro:Hessianoffdiag}
     H_{mn} = -\left( \frac{1}{|m - n|^\gamma} + \frac{1}{(m + n + 1)^\gamma} \right).
\end{equation}
Where $\tau \equiv \frac{2 \pi T}{g}$ is the dimensionless temperature, and $g$ is the bare coupling constant. Notice the remarkable absence of any parameters except dimensionless $\gamma$ and $\tau$, which allows for calculation of the transition temperature as a function of $\gamma$ only. The normal state is a minimum when all eigenvalues of the Hessian are positive. It becomes a saddle point when one of the eigenvalues changes sign. This is where the superconducting transition happens.

To analyze the eigenvalues numerically, we need to truncate the linear Hessian matrix to a finite $N\times N$ matrix. The size $N$ should be such that the Zeeman field overwhelms the interaction near the end of the chain as it does at large Matsubara frequencies in an infinite chain. This translates into the requirement $ N \gg \tau^{-\gamma}$ for $\gamma >1$ and $(N\tau)^\gamma \gg1$ for $\gamma<1$ \cite{emil2}. However, while this approach is both numerically and computationally practical, it presents two key subtleties that must be addressed before it can be reliably applied:

\begin{itemize}
    \item \textbf{Infinite-dimensional Hessian matrix:} The Hessian matrix in question is infinite-dimensional, with diagonal elements \( H_{nn} \to \infty \) as \( n \to \infty \) for any $\gamma >0, \tau>0$. This implies that the matrix \( H \) is neither compact nor bounded. Consequently, truncating it to a finite \( N \times N \) matrix is not trivially justified and must be done carefully.
    
    \item \textbf{Dependence on prior knowledge of the transition temperature:} Even if truncation were feasible, the conditions \( N \gg \tau^{-\gamma} \) for \( \gamma > 1 \) and \( (N\tau)^\gamma \gg 1 \) for \( \gamma < 1 \) require prior knowledge of the transition temperature \( \tau_c \), or at least its bounds. This creates a circular dependency: one needs to know \( \tau_c \) to determine the truncation size necessary to compute \( \tau_c \). While previous studies have estimated the transition temperature using alternative methods \cite{PhysRevB.99.144512,paper1,paper2,paper3,paper4,paper5,paper6}, a self-contained approach based on the spin chain mapping only would be more comprehensive.
\end{itemize}

These issues were recently addressed by the original authors of the spin chain mapping approach \cite{Kiessling2025,Kiessling2024b,Kiessling2024c}. They introduced a compact, self-adjoint operator \( \mathfrak{G}(\gamma) \), identifying the transition temperature as (see Eq.~(46) in \cite{Kiessling2025}):

\[
T_c = \frac{1}{2\pi} [\mathfrak{g}(\gamma)]^{\frac{1}{\gamma}},
\] where \( \mathfrak{g}(\gamma) \) is the largest eigenvalue of \( \mathfrak{G}(\gamma) \) and by setting the bare electron-boson coupling constant $g=1$. They introduced a variational method to derive lower bounds on this eigenvalue, and hence on the transition temperature, and used fixed-point theory to establish an upper bound.

While their results are mathematically rigorous, they are primarily tailored to a mathematically inclined audience. Moreover, the upper bound derived in \cite{Kiessling2025} is exceedingly high across a wide range of $\gamma$, suggesting clear potential for improvement. While a $1/\gamma$ power dependence eventually drives their upper bound to $1$ in the limit $\gamma \to \infty$, it fails to approach the true asymptotic behavior $\tau_c \sim (1.1843)^{1/\gamma}$ obtained in \cite{emil2,wang}. Instead, their bound scales asymptotically as $\tau_c \sim (5.2991)^{1/\gamma}$, rendering it quantitatively loose for finite and even large values of $\gamma$. Our work aims to address these two points. We build on their mathematical argument and show that it is possible to establish a truncation for the original Hessian matrix without the need for introducing a new matrix. We also show that, using a simple linear algebra framework and even when taking the limit $N \to \infty$, we can establish sharper upper bounds.

\subsection{Overview of our results}
The main results of this work can be summarized in the following points:
\begin{itemize}
    \item Building upon the foundational rigorous proofs developed in \cite{Kiessling2025}, we detailed for a physics audience how it is mathematically justified to apply truncation directly to the infinite unbounded Hessian matrix. This finite $N \times N$ truncation allows for the calculation of strictly increasing lower approximations to the transition temperature at which a negative eigenvalue appears.
    
    \item Increasing the size $N$ will increase the accuracy of the transition temperature calculation. We independently calculated the exact temperatures for the sizes $N=1,2,3,4$ which can be taken as progressing lower bounds approaching the actual temperature from below. The $N$-th bound is obtained as the largest solution of the equation $\det(H^{(N)}(\tau,\gamma))=0$ equation where $H^{(N)}(\tau,\gamma)$ is the $N \times N$ truncation of the Hessian matrix as a function of the dimensionless temperature $\tau$, and the parameter $\gamma$. These results independently confirm the lower bounds on $T_c$ obtained in \cite{Kiessling2025}.
    \item We establish a significantly improved upper bound on the transition temperature using Gershgorin circle theorem and by scaling the original Hessian matrix through a similarty transformation. This sharpens the bound on the eigenvalues of the Hessian in terms of the matrix entries, which enables us to establish a bound on the transition temperature $\tau_c$.
\end{itemize}

\begin{figure}
    \centering
    \includegraphics[width=1\linewidth]{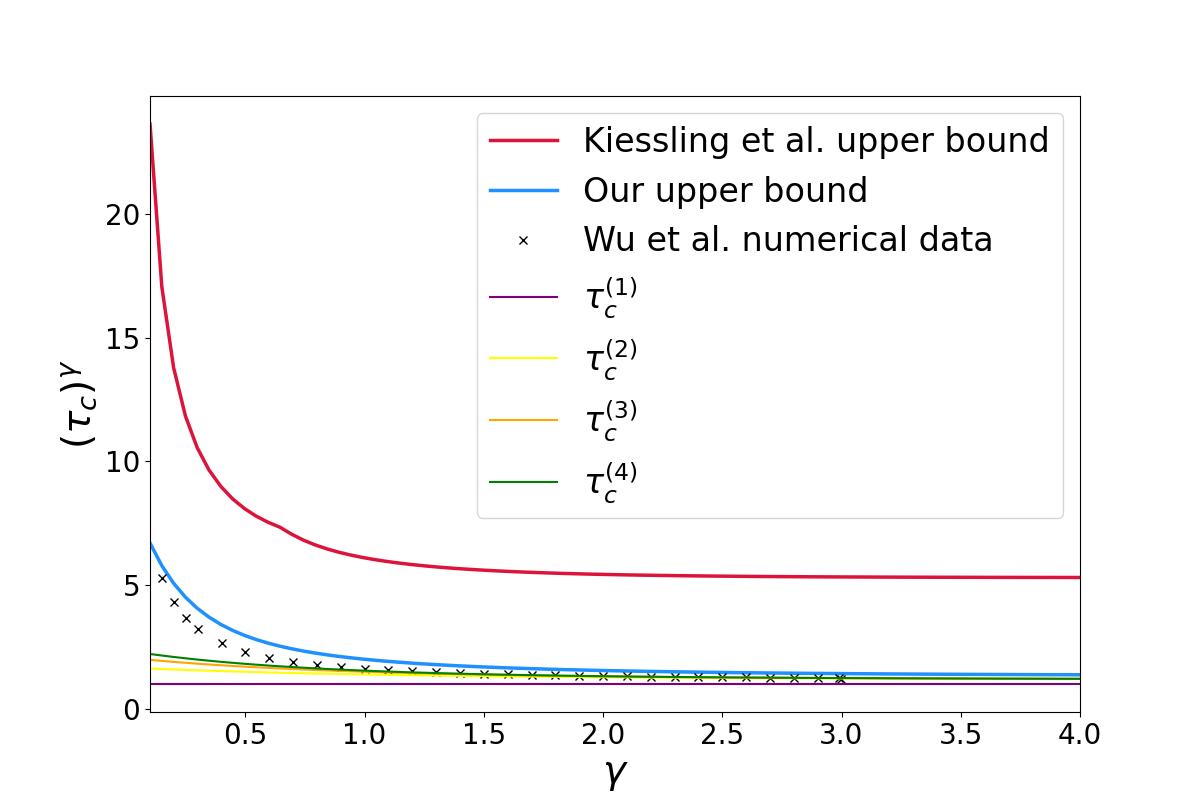}
    \caption{The first four lower bounds $\tau_{c}^{(1)},\tau_{c}^{(2)},\tau_{c}^{(3)},\tau_{c}^{(4)}$ as computed from Eq.s and(\ref{eq:Lower}),(\ref{tau0}),and (\ref{tau1}), and our upper bound calculated from Eq (\ref{eq:ourbound}) in comparison with the numerical data from \cite{PhysRevB.99.144512} and the upper bound obtained by Kiessling et al \cite{Kiessling2025}. Note that we are plotting $\tau^\gamma$ instead of $\tau$ vs $\gamma$.}
    \label{fig:full}
\end{figure}

Our results are summarized in Fig.~\ref {fig:full}.

\subsection{Structure of the paper}
The paper is structured as follows. In section \ref{sec:gammamodel} we review the $\gamma$-model and its mapping to a classical spin chain. We derive the gap equation and the Hessian matrix needed to do our analysis. In section \ref{sec:math}, we present the mathematical argument for the truncation of the Hessian matrix based on the framework established in \cite{Kiessling2025}. In section \ref{sec:lower}, we reproduce the results from \cite{Kiessling2025} for the lower bounds on the transition temperature using the Rayleigh-Ritz variational method directly on the truncated Hessian. Section \ref{sec:upper} contains our main result. We use the Gershgorin Circle theorem to establish an upper bound for the transition temperature. The summary and conclusions are presented in section \ref{sec:summary}.

\section{ \label{sec:gammamodel} The $\gamma$-Model}
Pairing near a Quantum Critical Point (QCP) arises from a mechanism fundamentally different from that of conventional BCS theory \cite{wang,PhysRevB.99.144512}. In this regime, the effective dynamic electron-electron interaction \( V(\mathbf{q}, \Omega) \) is mediated by a critical collective bosonic mode. The condensation of this boson at the QCP provides strong attraction in one or more pairing channels. Simultaneously, it induces non-Fermi liquid (NFL) behavior in the normal state. These two tendencies—pairing and NFL behavior—compete, and determining which dominates requires analyzing a set of coupled nonlinear integral equations for the fermionic self-energy \( \Sigma(\mathbf{k}, \omega) \) and the gap function \( \Delta(\mathbf{k}, \omega) \), where \( \mathbf{k} \) and \( -\mathbf{k} \) are the momenta of paired fermions and \( \omega \) is the Matsubara frequency.

We consider a class of models in which the collective bosons are slow modes compared to the dressed fermions, due to various physical reasons. In this limit, the self-energy and pairing vertex can be approximated by their values on the Fermi surface, similar to the treatment in the standard Eliashberg theory \cite{eliash1,eliash2,Eliashberg2}.

A key feature of the \( \gamma \)-model is that the effective dynamical interaction takes the form \( V(\Omega_m) \propto 1/|\Omega_m|^\gamma \), where \( \Omega_m \) is the transferred bosonic Matsubara frequency. Near the QCP, the fermion-boson interaction becomes infinitely strong. In this limit, the only remaining energy scale in the problem is the bare fermion-boson coupling strength, denoted by \( g \) in this paper. As a result, the system becomes universal, with its behavior governed solely by the nature of the bosonic mode, encapsulated in the dimensionless parameter \( \gamma \). This universality is what defines the \( \gamma \)-model.

\subsection{Generalized Eliashberg Equations}

Near a Quantum Critical Point (QCP), the bosonic gap vanishes, and the bosonic propagator takes the form:
\begin{equation}
    \chi(\omega) \sim |\omega|^{-\gamma},
\end{equation}
where \( \gamma > 0 \) is a dimensionless parameter that depends on the nature of the QCP.

The generalized Eliashberg equations for arbitrary \( \gamma \) can be derived in the Matsubara formalism. They are given by \cite{abanovnormal,paper1,paper2,paper3,paper4,paper5,paper6}: (the Matsubara frequencies $\omega_m=\pi T(2m+1)$, the anomalous vertex $\Phi_n\equiv\Phi(\omega_n)$, and the electron self-energy ${\Sigma}_m\equiv{\Sigma}(\omega_m)$)
\begin{equation}\label{eq:gammaEliashberg1}
    \Phi_m = g^\gamma \pi T \sum_{m' \neq m} \frac{\Phi_{m'}}{\sqrt{ {\Sigma}^2_{m'} + \Phi^2_{m'}}}  \frac{1}{|\omega_m - \omega_{m'}|^\gamma},
\end{equation}
and
\begin{equation}
{\Sigma}_m = \omega_m 
+ g^\gamma \pi T \sum_{m' \neq m} 
\frac{{\Sigma}_{m'}}{\sqrt{{\Sigma}^2_{m'} + \Phi^2_{m'}}} 
\frac{1}{|\omega_m - \omega_{m'}|^\gamma}.
\label{eq:gammaEliashberg2}
\end{equation}

The superconducting gap function \( \Delta_m \) is defined as:
\begin{equation}
    \Delta_m = \omega_m  \frac{\Phi_m}{ {\Sigma}_m}.
\end{equation}
Substituting this into the equations above, we obtain a single self-consistent equation for the gap function:
\begin{equation}\label{eq:gammaGap}
    \Delta_m =  g^\gamma \pi T\sum_{m' \neq m} \frac{\Delta_{m'} - \Delta_m \frac{\omega_{m'}}{\omega_m}}{\sqrt{\omega_{m'}^2 + \Delta^2_{m'}}}  \frac{1}{|\omega_m - \omega_{m'}|^\gamma}.
\end{equation}

This equation involves only the gap function \( \Delta(\omega_m) \), but it appears on both sides, making it a nonlinear summation equation that must be solved self-consistently.

In the right-hand side of equations \cref{eq:gammaEliashberg1,eq:gammaEliashberg2,eq:gammaGap}, we have excluded the resonance term ($m=m'$) for the self-energy, the pairing vertex, and the gap function. This procedure is legitimate because such terms come from thermal fluctuations, which act as non-magnetic impurities. In an $s$-wave superconductor, the ($m=m'$)
contributions to the self energy and the pairing vertex cancel out exactly, by Anderson’s theorem \cite{Millis1988,Abanov2008}. Notably, while Anderson’s Theorem traditionally only protects $s$-wave states, it was demonstrated in \cite{wang} that in the $\gamma$-model, the resonance term cancellation extends to non-$s$-wave spin-singlet states. This is a consequence of the local (momentum-independent) nature of the Eliashberg approximation, where the static fluctuations renormalize the single-particle mass and the pairing vertex identically, effectively bypassing the pair-breaking typically associated with all-angle impurity scattering in $d$-wave systems. We refer the reader to the Supplemental Material of \cite{wang} for the complete theoretical arguments and supporting calculations for both cases.

\subsection{Mapping to the Classical Spin Chain}

In \cite{emil1}, the Migdal–Eliashberg theory of superconductivity was mapped onto a classical Heisenberg spin chain in a Zeeman magnetic field, where the lattice sites correspond to Fermionic Matsubara frequencies \( \omega_m \). In terms of classical spin variables \( \mathbf{S}_m \), the Holstein Hamiltonian \( H_s \) can be written as:
\begin{eqnarray}
H_s(\omega_m) = -2\pi \sum_m \omega_m S^z_m 
- g^2 \pi^2 T \!\!\sum_{m' \neq m} 
\frac{\mathbf{S}_m \cdot \mathbf{S}_{m'} - 1}{|\omega_m - \omega_{m'}|^\gamma},
\end{eqnarray}
where \( T \) is the temperature and \( g \) is the electron-phonon coupling constant.

Introducing the frequency-dependent gap function \( \Delta_m \equiv \Delta(\omega_m) \) through the spin components:
\begin{equation}
    S^z_m = \frac{\omega_m}{\sqrt{\omega_m^2 + |\Delta_m|^2}}, \quad S^+_m = \frac{\Delta_m}{\sqrt{\omega_m^2 + |\Delta_m|^2}},
\end{equation}
where \( S^+_m = S^x_m + i S^y_m \).

It was shown in \cite{emil1,emil2} that, with an appropriate choice of the overall phase, the gap function at the free energy minimum satisfies the following properties:
\begin{itemize}
    \item \( \Delta_m \geq 0 \) for all \( \omega_m \),
    \item \( \Delta_m \) is an even function of \( \omega_m \),
    \item \( \Delta_m \to 0 \) as \( \omega_m \to \pm \infty \).
\end{itemize}
These properties also hold in the zero-temperature limit \( T = 0 \), where \( \omega_m \) becomes a continuous variable \( \omega \).

From these properties, it follows that \( S^y_m = 0 \) and \( S^x_m \geq 0 \). Let \( \theta_m \) be the angle between \( \mathbf{S}_m \) and the \( z \)-axis. Then:
\begin{equation*}
    S^z_m = \cos \theta_m, \quad S^x_m = \sin \theta_m, \quad 0 \leq \theta_m \leq \pi.
\end{equation*}
The gap function can then be expressed as:
\begin{equation}
    \Delta_m = \omega_m \tan \theta_m,
\end{equation}
subject to the boundary conditions:
\begin{equation}\label{boundary}
    \theta(-\omega_m) = \pi - \theta(\omega_m), \quad \lim_{\omega_m \to \pm \infty} \theta_m = \frac{\pi}{2}\mp \frac{\pi}{2}.
\end{equation}

In terms of the angles \( \theta_m \), the spin chain Hamiltonian becomes \cite{emil2}:
\begin{eqnarray}
&&
H_s(\theta_m) = -2\pi \sum_m \omega_m \cos \theta_m 
\nonumber\\
&&
- g^\gamma \pi^2 T \sum_{m' \neq m} \frac{\cos(\theta_m - \theta_{m'}) - 1}{|\omega_m - \omega_{m'}|^\gamma}.
\label{eq:chainH}
\end{eqnarray}

Differentiating with respect to \( \theta_m \), we obtain the equation for stationary points of the \( \gamma \)-model:
\begin{equation} \label{eq:spingap}
    \omega_m \sin \theta_m = g^\gamma \pi T \sum_{m' \neq m} \frac{\sin(\theta_m - \theta_{m'})}{|\omega_m - \omega_{m'}|^\gamma}.
\end{equation}

Comparing this with Eq.~\eqref{eq:gammaGap}, we see that this is indeed the superconducting gap equation expressed in terms of the 
angles $\theta_m$.

\subsection{Boundary Conditions}

From Eq.~\eqref{eq:spingap}, we can rewrite the equations as: 
\begin{equation}\label{eq:boundary}
     \sin{\theta_m} =\frac{1}{(2m+1)} \left(\frac{g}{2\pi T}\right)^\gamma \sum_{m' \neq m} \frac{\sin(\theta_m -\theta_{m'})}{|m-m'|^\gamma}.
\end{equation}
For $\gamma > 1$, the summation on the right-hand side is strictly bounded, since $|\sin(\theta_m -\theta_{m'})| \leq 1$ and $\sum_{n=1}^\infty \frac{1}{n^\gamma}$ converges. Because the sum is bounded, the $\frac{1}{2m+1}$ prefactor drives the right-hand side to zero as $m \to \infty$, implying that $\theta_m \to 0$.

For the regime $0 < \gamma \leq 1$, the analysis is more subtle because absolute convergence is no longer guaranteed. Looking at the structure of Eq.~\eqref{eq:boundary}, the boundedness of the left-hand side ($|\sin\theta_m| \leq 1$) restricts the summation on the right-hand side from growing faster than $O(m)$ as $m \to \infty$. For a solution to exist where $\theta_m$ does not vanish in the large-$m$ limit, the summation would have to diverge at a rate that precisely counteracts the $\frac{1}{2m+1}$ prefactor. Physically, any such non-vanishing asymptotic behavior would require high-energy modes to remain coupled to the pairing condensate, causing the gap function to grow unboundedly at high frequencies. This violates standard UV decoupling and results in an unphysical divergence of the superconducting condensation energy. Therefore, the boundary condition $\lim_{m \to \infty} \theta_m = 0$ emerges as the unique physically admissible and self-consistent asymptotic behavior for the gap function, a choice further validated by the agreement between our analytical results and prior numerical data.

\subsection{The Free Energy Functional and the Hessian Matrix}

From the definition of the Fermionic Matsubara frequency \( \omega_m = (2m+1)\pi T \), the Hamiltonian exhibits the following symmetries:
\begin{equation}
    \omega_{-m-1} = -\omega_m, \quad \theta_{-m-1} = \pi - \theta_m.
\end{equation}
These symmetries allow us to express the Hamiltonian using only non-negative indices \( m \geq 0 \):
\begin{equation}
\begin{aligned}
&H_s = -4\pi \sum_{m=0}^\infty \omega_m \cos \theta_m \\
&- 2g^{\gamma} \pi^2 T \sum_{\substack{n,m=0 \\ n \neq m}}^\infty 
\frac{\cos(\theta_m - \theta_n) - 1}{|\omega_m - \omega_n|^\gamma} \\
&+2g^{\gamma} \pi^2 T\sum_{n,m=0}^\infty 
\frac{\cos(\theta_m + \theta_n) + 1}{(\omega_m + \omega_n)^\gamma} .
\end{aligned}
\end{equation}

We introduce the dimensionless temperature:
\begin{equation}\label{eq:tau}
    \tau = \frac{2\pi T}{g}.
\end{equation}

The free energy density \( f \) for a given field configuration is:
\begin{equation*}
    f = \nu_0 T H_s,
\end{equation*}
where \( \nu_0 \) is the density of states at the Fermi level. We define the total free energy functional as:
\begin{equation}
    F(\{\theta_m\}) \equiv \frac{2f(\{\theta_m\})}{g^2 \nu_0},
\end{equation}
which, in terms of the angles \( \theta_m \) and the dimensionless temperature \( \tau \), becomes:
\begin{equation}\label{eq:F}
\begin{aligned}
&F(\{\theta_m\}) = -2\tau^2 \sum_{m=0}^\infty (2m+1) \cos \theta_m \\
&- \tau^{2-\gamma} \sum_{\substack{n,m=0 \\ n \neq m}}^\infty 
\frac{\cos(\theta_m - \theta_n) - 1}{|\omega_m - \omega_n|^\gamma} \\
&+\tau^{2-\gamma} \sum_{n,m=0}^\infty 
\frac{\cos(\theta_m + \theta_n) + 1}{(\omega_m + \omega_n)^\gamma} .
\end{aligned}
\end{equation}

To obtain the Hessian matrix at the normal state (i.e., \( \theta_n = 0 \) for all \( n \)), we compute the second derivative of the free energy functional \eqref{eq:F}. The resulting Hessian matrix \( \tau^{2-\gamma}\tilde{H} \) has elements:
\begin{eqnarray}
&&\tau^{2-\gamma}\tilde{H}_{nn} = 2\tau^{\gamma} (2n+1) \cos \theta_n 
- 2 \frac{\cos(2\theta_n)}{(2n+1)^\gamma} \\
&& + 2 \Bigg( 
\sum_{\substack{m=0 \\ m \neq n}}^\infty \frac{\cos(\theta_m - \theta_n)}{|m - n|^\gamma} 
- \sum_{m=0}^\infty \frac{\cos(\theta_m + \theta_n)}{(m + n + 1)^\gamma} \Bigg).
\nonumber
\end{eqnarray}
and for \( m \neq n \):
\begin{equation}
   \tau^{2-\gamma} \tilde{H}_{mn} = -2 \left(\frac{\cos(\theta_m - \theta_n)}{|m - n|^\gamma} +\frac{\cos(\theta_m + \theta_n)}{(m + n + 1)^\gamma} \right).
\end{equation}

Expanding the free energy around the normal state yields:
\begin{equation}
    \delta F = F - F_N = 2\tau^{2-\gamma} \sum_{n,m=0}^\infty \theta_n H_{nm} \theta_m,
\end{equation}
where \( F_N \) is the free energy of the normal state, and \( H \) is the linearized Hessian matrix evaluated at \( \theta_n = 0 \). The elements of \( H \) are given by:
\begin{equation}\label{eq:Hessiandiag}
     H_{nn} = (2n+1)\tau^\gamma + 2 \sum_{m=1}^{n} \frac{1}{m^\gamma} - \frac{1}{(2n+1)^\gamma},
\end{equation}
and for \( m \neq n \):
\begin{equation}\label{eq:Hessianoffdiag}
     H_{mn} = -\left( \frac{1}{|m - n|^\gamma} + \frac{1}{(m + n + 1)^\gamma} \right).
\end{equation}
These are the Hessian matrix elements we refered to in the equations \eqref{eq:intro:Hessiandiag} and \eqref{eq:intro:Hessianoffdiag}.

\section{\label{sec:math} Truncation of the Hessian Matrix}

In the functional analysis of linear operators on infinite-dimensional Hilbert spaces, the reliability of numerical truncation (the finite section method) is deeply tied to the compactness and boundedness properties of the operator. While bounded operators feature a finite operator norm and a bounded spectral radius, many operators defined on infinite-dimensional spaces—including differential operators and infinite matrices with asymptotically growing elements—are unbounded, possessing eigenvalues that scale to infinity \cite{Kato1995, Maddox1970}. A more fundamental topological distinction concerns compactness. A compact operator maps bounded sequences to sequences containing a convergent subsequence, a property which guarantees a purely discrete point spectrum accumulating only at zero. Although unbounded operators lack compactness, those defined on bounded domains (or sequence spaces with appropriate decay) typically possess a compact resolvent, defined as $R_z = (A - zI)^{-1}$. As established by the spectral theorem, if the operator $A$ is self-adjoint, the compactness of its resolvent ensures a purely discrete spectrum accumulating only at infinity; for typical physical systems, this spectrum is additionally bounded from below \cite{ReedSimon1980}.

Consequently, the convergence of finite-dimensional approximations (such as matrix truncations) is contingent upon preserving the spectral topology of the underlying operator. For operators with a compact resolvent, such as the stiffness matrix of a fixed-boundary elastic string or a confining potential well, truncation is mathematically well-posed; the Min-Max principle guarantees that the eigenvalues of the finite projection converge variationally to the exact discrete eigenvalues of the operator \cite{Courant1953}. Conversely, for operators exhibiting a \emph{continuous spectrum}, characteristic of translationally invariant systems like infinite lattices, free fields, or acoustic scattering, finite-dimensional truncation is topologically ill-posed. In these regimes, the discrete eigenvalues of the truncated matrix attempt to approximate a continuous spectral density, resulting in "spectral pollution" or spurious eigenvalues that depend on the truncation dimension $N$ rather than physical parameters \cite{Bottcher1999}.

To this end, we are interested in showing that, at least for the search for negative points in the spectrum, truncation can still work and be mathematically justified. We now explain the analytical methods developed in \cite{Kiessling2025} that establish a robust justification for the Hessian matrix truncation that is valid for any $\gamma > 0$. The authors in \cite{Kiessling2025} showed that the transition temperature $T_c$ corresponds to the largest eigenvalue $\mathfrak{g(\gamma)}$ of a compact operator $\mathfrak{G}$. We show that this is equivalent to the standard procedure of finding the temperature $T_c$ at which the lowest eigenvalue of the Hessian matrix crosses zero. 

The stability of the normal state is determined by the second-order variation of the free energy, which defines a quadratic form, or Hessian operator, $K^{(2)}$. Its action on the sequence $\Theta = (\theta_n)$ is given by:
\begin{align}
    &K^{(2)}(\Theta)_\gamma = \sum_{n,m} \theta_n H_{nm} \theta_m \nonumber \\ 
    &= \sum_{n=0}^{\infty} \left[ (2n+1)\tau^\gamma + 2 \sum_{k=1}^{n} \frac{1}{k^\gamma} \right] \theta_n^2 \notag \\
    & -\sum_{\substack{m=0\\n\not=m}}^{\infty} \theta_n \left[\frac{1}{|n - m|^\gamma} + \frac{1}{(n + m + 1)^\gamma} \right] \theta_m
\end{align}
This functional is precisely functional Eq.~(38) in~ \cite{Kiessling2025}, except for an overall factor of $\frac{1}{2}\tau^\gamma$. It is well-defined on the Hilbert space $\mathcal{H}$ of real sequences that satisfy the condition $||\Theta||^2_\mathcal{H} := \sum_{n=0}^{\infty} (2n+1) \theta_n^2 < \infty$. While direct analysis in $\mathcal{H}$ is challenging due to the weighted norm, we can simplify the problem by mapping it to the standard Hilbert space $\ell^2(\mathbb{N}_0)$ of square-summable sequences. This is achieved through the transformation $\xi_n := \sqrt{2n+1} \, \theta_n$. In terms of the new sequence $\Xi = (\xi_n)$, the quadratic form, denoted $Q_\gamma(\Xi)$, becomes:
\begin{align}
    &Q_\gamma(\Xi)\! :=\! K^{(2)}(\Theta)_\gamma \! 
    = \! \sum_{n=0}^{\infty} \left[ \tau^\gamma\! +\! \frac{2}{2n+1}\sum_{k=1}^{n} \frac{1}{k^\gamma} \right] \xi_n^2 \\
    &-\!\sum_{\substack{m=0\\n\not=m}}^{\infty} \frac{\xi_n}{\sqrt{2n+1}} \left[\frac{1}{|n - m|^\gamma} + \frac{1}{(n + m + 1)^\gamma} \right] \frac{\xi_m}{\sqrt{2m+1}} \notag
\end{align}
This new functional is precisely functional Eq.~(40) in~ \cite{Kiessling2025}, except for an overall factor of $\tau^\gamma$. A superconducting instability corresponds to $K^{(2)}_\gamma < 0$. Since this transformation preserves the sign of the functional, this condition is equivalent to $Q_\gamma < 0$. We can therefore analyze the stability of the more tractable operator associated with $Q_\gamma$. This functional can be written as $Q_\gamma(\Xi ) = \bra{\Xi}\tau^\gamma\tilde{X}\ket{\Xi}$, where $\tilde{X} =  \mathcal{I} - \tau^{-\gamma}\mathfrak{G}$, $\mathcal{I}$ is the identity operator, and the operator $\mathfrak{G}$ is expressed as $\mathfrak{G}= -\mathfrak{G}_1 + \mathfrak{G}_2 +\mathfrak{G}_3$, whose components are given by:
\begin{align}
    &(\mathfrak{G}_1(\gamma) \Xi)_n = \left[ \frac{2}{2n + 1} \sum_{k=1}^{n} \frac{1}{k^\gamma} \right] \xi_n, \\
    &(\mathfrak{G}_2(\gamma) \Xi)_n \!=\!\!\sum_{m \ne n} 
    \frac{1}{\sqrt{2n + 1}} \frac{1}{|n - m|^\gamma} \frac{\xi_m}{\sqrt{2m + 1}} 
    , \\
    &(\mathfrak{G}_3(\gamma) \Xi)_n\! =\!\! \sum_{m=0}^{\infty} 
    \frac{1}{\sqrt{2n + 1}} \frac{1}{(n + m + 1)^\gamma} \frac{\xi_m}{\sqrt{2m + 1}} 
    .
\end{align}
The key insight from \cite{Kiessling2025} is that each of the operators $\mathfrak{G}_i$ for $i \in \{1,2,3\}$ is a \emph{compact operator} on $\ell^2$. This property is crucial, as it implies that their sum, $\mathfrak{G}$, is also compact. A compact operator is, in essence, an infinite-dimensional matrix whose influence effectively vanishes for large indices, meaning it can be arbitrarily well-approximated by a finite-rank matrix. The full operator $\tilde{X}$ is the sum of a simple self-adjoint operator, $\mathcal{I}$, and the compact operator $-\tau^{-\gamma}\mathfrak{G}$. By Weyl's theorem \cite{Weyl1909} on the essential spectrum, the addition of a compact operator does not alter the essential spectrum of a self-adjoint operator. The essential spectrum of $ \mathcal{I}$ is simply the set $\{1\}$, and thus $\sigma_{\text{ess}}(\tilde{X}) = \{1\}$.

Since the full spectrum is the union of the essential and discrete spectra, $\sigma(\tilde{X}) = \sigma_{\text{ess}}(\tilde{X}) \cup \sigma_{\text{disc}}(\tilde{X})$, any eigenvalue that could cross zero to signal an instability must belong to the \emph{discrete spectrum}. The discrete spectrum is generated by the compact part of the operator, $-\tau^{-\gamma}\mathfrak{G}$. This provides the rigorous basis for the approximation: the search for an instability is equivalent to finding the largest discrete eigenvalue of $\tau^{-\gamma}\mathfrak{G}$ and determining when it exceeds $1$. Because these critical eigenvalues are in the discrete spectrum of a compact operator, they can be found by truncating $\tau^{-\gamma}\mathfrak{G}$ to a finite $N \times N$ matrix, $\mathfrak{G}_N$, and calculating its eigenvalues numerically. We also notice that finding $\tau$ for a finite truncation of this compact operator has an eigenvalue greater than $1$ is equivalent to a finite truncation of $\tilde{X}$ has a negative eigenvalue. Since $\tilde{X}$ is bounded, the only issue we need to worry about is the essential spectrum, which is only $\{1\}$, thus contains no negative values.  Finally, we note that the original Hessian operator $H$ is related to $\tilde{X}$ by the congruence transformation $H = \mathfrak{D} \tilde{X} \mathfrak{D}$, where $\mathfrak{D}$ is the invertible diagonal matrix defined in Eq ~(39) in \cite{Kiessling2025}, with entries $\mathfrak{D}_{nn} = \sqrt{2n+1}$. From Sylvester's law of inertia, this transformation preserves the signs of the eigenvalues since $\mathfrak{D}$ is a diagonal positive definite matrix. Therefore, $H$ has a zero or negative eigenvalue if and only if $\tilde{X}$ does. This completes the justification for using a finite Hessian matrix truncation to find the stability threshold for any $\gamma > 0$.

We emphasize that while the compactness of $\mathfrak{G}$ was established by Kiessling \emph{et al.} \cite{Kiessling2025} to formulate a well-posed eigenvalue problem, the application of Sylvester’s Law of Inertia provides a critical extension to standard numerical practices. In routine numerical analysis, one directly truncates the unbounded Hessian $H$ to search for a zero-crossing, a procedure that is generally susceptible to spectral pollution. By showing that the exact congruence between $H$ and $\tilde{X}$ holds at any finite truncation dimension $N$, we demonstrate that the discrete numerical search for $\tau_c$ using the unbounded Hessian is topologically protected from spurious continuous spectra and strictly equivalent to evaluating the eigenvalue problem of the compact operator introduced in \cite{Kiessling2025}. This bridges the rigorous operator formalism of  \cite{Kiessling2025} with standard numerical routines, proving that a finite matrix truncation of the unbounded Hessian is topologically protected and mathematically valid for locating the stability threshold for any $\gamma > 0$.

Finally, to demonstrate that this instability, specifically, the smallest eigenvalue of the Hessian crossing zero to become negative, actually corresponds to a physical superconducting instability, we must confirm two conditions: the uniqueness (non-degeneracy) of this lowest eigenvalue, and that the components of its corresponding eigenvector do not change sign. These two properties follow directly from the Perron-Frobenius theorem. To see this, consider an $N \times N$ truncation of the Hessian matrix, denoted as $H^{(N)}$. We define a shifted matrix $M = sI - H^{(N)}$, where $I$ is the identity matrix and $s$ is a large, positive constant chosen such that all diagonal elements $M_{nn}$ are strictly positive. As shown in Section~\ref{sec:upper}, $\tau_c$ possesses a finite upper bound (except in the limit $\gamma \to 0$), ensuring that a finite constant $s$ can always be selected. Given that the off-diagonal entries of $H^{(N)}$ are strictly negative, the matrix $M$ consists entirely of positive entries. By the Perron-Frobenius theorem, $M$ is guaranteed to have a unique, non-degenerate largest eigenvalue, and its corresponding eigenvector features strictly positive components (up to a trivial overall sign). Because the eigenvalues of the two matrices are related via $\lambda_H = s - \lambda_M$, the maximum eigenvalue of $M$ corresponds precisely to the unique, non-degenerate minimum eigenvalue of $H^{(N)}$. Furthermore, since $H^{(N)}$ and $M$ share the exact same eigenvectors, the eigenvector associated with the lowest eigenvalue of $H^{(N)}$ inherits these strictly positive components.

\section{\label{sec:lower} Lower Bounds}
In this section, we start analyzing the bounds on the transition temperature $\tau_c$ by reproducing the results obtained in \cite{Kiessling2025} directly from the truncated Hessian matrix. To this end, we invoke the Rayleigh-Ritz variational principle \cite{Courant1953, Horn2012}, which characterizes the eigenvalues of a symmetric matrix through the optimization of the Rayleigh quotient. For any finite $N \times N$ truncation of the Hessian matrix, denoted as $H^{(N)}$, the smallest eigenvalue $\lambda_1^{(N)}$ is determined by minimizing the Rayleigh quotient over the continuous domain of all non-zero vectors $x \in \mathbb{R}^N$:$$\lambda_1^{(N)} = \min_{x \neq 0} \frac{x^T H^{(N)} x}{x^T x}$$The smaller truncated matrix $H^{(N-1)}$ corresponds to evaluating this exact same quadratic form, but restricted to a smaller subspace. Specifically, any vector $v \in \mathbb{R}^{N-1}$ can be trivially embedded into the larger continuous space $\mathbb{R}^N$ by appending a zero to its final component, $\tilde{v} = (v_0, \dots, v_{N-2}, 0)^T$. Therefore, finding the lowest eigenvalue of $H^{(N-1)}$ is mathematically equivalent to minimizing the Rayleigh quotient of $H^{(N)}$ subject to the strict constraint that the $(N-1)$-th component of the trial vector is zero \footnote{Note that indexing starts from $0$, i.e $i=0,1,\cdots$, consistent with the Hessian matrix indexing.}. A fundamental property of optimization is that minima of continuous functions on bounded, closed, finite-dimensional domains can at most increase when one shrinks the domain, yielding the inequality:
$$\lambda_1^{(N)} \leq \lambda_1^{(N-1)}$$

Furthermore, the specific structure of $H^{(N)}$ strictly forbids the equality condition. As established, the shifted matrix $M = sI - H^{(N)}$ contains strictly positive entries, allowing us to invoke the Perron-Frobenius theorem. This guarantees that $\lambda_1^{(N)}$ is non-degenerate and that the unique minimizing eigenvector $x \in \mathbb{R}^N$ possesses strictly positive components ($x_i > 0$ for all $i=0,\dots,N-1$). Because the restricted trial vector $\tilde{v}$ defined above explicitly requires $\tilde{v}_{N-1} = 0$, it cannot be collinear with the true eigenvector $x$. Evaluating the Rayleigh quotient  at $\tilde{v}$ must yield a strictly larger value:$$\frac{\tilde{v}^T H^{(N)} \tilde{v}}{\tilde{v}^T \tilde{v}} > \lambda_1^{(N)}$$Taking the minimum of the left-hand side over all non-zero $v \in \mathbb{R}^{N-1}$ directly upgrades the weak inequality to a strict one,
$$\lambda_1^{(N-1)} > \lambda_1^{(N)}$$

By successively applying this variational argument as the truncation size increases, similar to the argument invoked in \cite{Kiessling2025}'s Eqs~(52-53), we obtain a strictly decreasing chain of bounds for the lowest eigenvalue:

\begin{equation}\label{interlacing}\lambda = \lambda_1^{(\infty)} < \cdots < \lambda_1^{(4)} < \lambda_1^{(3)} < \lambda_1^{(2)} < \lambda_1^{(1)}\end{equation}Thus, if $\lambda_1^{(k)} < 0$ for any finite truncation dimension $k$, the variational principle guarantees that the true minimum over the full, unconstrained parameter space cannot exceed this value. This provides a mathematically sufficient condition that the lowest eigenvalue of the full infinite matrix is also negative ($\lambda < 0$), rigorously confirming the onset of the instability.
\subsection{The sequence for $\tau_c^{(N)}$}
Expanding the matrix size strictly lowers its minimum eigenvalue, meaning the larger $(N+1)\times(N+1)$ matrix evaluated at the previous $\tau^{(N)}_c$ will have a negative eigenvalue. Because the diagonal elements of the Hessian strictly increase with $\tau$, the only mathematical way to push this lowest eigenvalue back up to exactly zero is to strictly increase $\tau$. This yields a strictly increasing sequence of transition temperatures $\tau_c^{(N)}$ as: 
\begin{equation}\label{tempsequence}
    \tau_c^{(1)} < \tau_c^{(2)} < \tau_c^{(3)} < \cdots < \tau_c^{(\infty)} = \tau_c,
\end{equation}
where $\tau_c$ here is understood to be the exact transition temperature obtained from the full untruncated Hessian matrix. The sequence (\ref{tempsequence}) is equivalent to the sequence in Eqs ~(15-16) in \cite{Kiessling2025}.

For convenience, we introduce the substitution $z=\tau^\gamma$ in Eq ~ \ref{eq:Hessiandiag}. The condition that any eigenvalue of the $N\times N$ Hessian matrix is zero is exactly the equation: 
\begin{equation} \label{eq:Lower}
    \det (H^{(N)}(z,\gamma)) =0
\end{equation}
This yields an $N^{th}$ degree polynomial in $z$, with at most $N$ distinct roots. The physical instability threshold, where the lowest eigenvalue vanishes ($\lambda_1^{(N)} = 0$), uniquely corresponds to the largest of these roots, $z_{max}$. 

To see this, notice that $z$ appears exclusively on the diagonal of $H^{(N)}$ with strictly positive coefficients, the matrix derivative $\frac{\partial H^{(N)}}{\partial z}$ is diagonal and positive-definite. By the Hellmann-Feynman theorem, every eigenvalue $\lambda_k(z)$ is a strictly increasing function of $z$. Furthermore, for any $z$, the spectrum is ordered (keeping in mind that the lowest eigenvalue is non-degenerate): $\lambda_1(z) < \lambda_2(z) \leq \dots \leq \lambda_N(z)$. Let $z_{max}$ be the largest root of the polynomial. By definition, at least one eigenvalue must vanish at this point; suppose it is $\lambda_k(z_{max}) = 0$. The spectral ordering then dictates that the lowest eigenvalue must satisfy:$$\lambda_1(z_{max}) \leq \lambda_k(z_{max}) = 0$$If we assume the strict inequality $\lambda_1(z_{max}) < 0$, the strict monotonicity of $\lambda_1(z)$ requires that it must eventually cross zero at some larger value $z^* > z_{max}$. However, the condition $\lambda_1(z^*) = 0$  requires $\det(H^{(N)}(z^*)) = 0$, which makes $z^*$ a valid root of the polynomial strictly greater than $z_{max}$. This contradicts the fundamental premise that $z_{max}$ is the largest root. We must therefore conclude that $\lambda_1(z_{max}) = 0$, proving that the lowest eigenvalue vanishes exactly at the largest root of the determinant. This means that we can compute $\tau_c^{(N)}$ directly as the largest root of Eq ~(\ref{eq:Lower}).

This is, in fact, precisely the result of \cite{Kiessling2025} obtained directly from the Hessian matrix. To see this equivalence, remember that the Hessian matrix at any truncation size $N$ is congruent to the matrix $\tilde{X}$ introduced in the previous section. The equation (\ref{eq:Lower}) can be rewritten as: 
\begin{equation*}
    \det(H^{(N)}) = \det(\mathfrak{D} \tilde{X} \mathfrak{D}) = 0
\end{equation*}
and since $\mathfrak{D}$ is nonsingular, this reduces to 
\begin{equation*}
  \det(\tilde{X}) = \det(I-z^{-1}\mathfrak{G})=\det(zI-\mathfrak{G})=0,
\end{equation*}
which is the characteristic equation of the matrix $\mathfrak{G}$. thus finding the largest root, $z_{max}$, of Eq. (\ref{eq:Lower}) is precisely calculating the largest eigenvalue of the matrix $\mathfrak{G}$.

\subsubsection{The bound $\tau_c^{(1)}$}

The matrix $H^{(1)}$ consists of a single element, which is also its only eigenvalue. Therefore,
\begin{equation}
    \lambda_1^{(1)} < 0 \quad \Rightarrow \quad \tau^\gamma - 1 < 0 \quad \Rightarrow \quad \tau < 1,
\end{equation}
which yields the bound
\begin{equation}\label{tau0}
    \tau_{c}^{(1)} = 1.
\end{equation}

\subsubsection{The Bound $\tau_{c}^{(2)}$}

The $2 \times 2$ matrix $H^{(2)}$ is given by
\begin{equation}
    H^{(2)} =\begin{bmatrix}
        H_{00} & H_{01} \\
        H_{10} & H_{11}
    \end{bmatrix} = \begin{bmatrix}
        z - 1 & -a \\
        -a & 3z + 2 - b
    \end{bmatrix},
\end{equation}
where \( a = (1 + \frac{1}{2^\gamma}) \) and \( b = \frac{1}{3^\gamma} \).

Equation (\ref{eq:Lower}) now reads: 
\begin{equation*}
    3z^2 -(b+1)z-(a^2-b+2)=0.
\end{equation*}
One can easily solve for the largest root using the standard quadratic formula, thus obtaining
\begin{equation}\label{tau1}
\left(\tau_{c}^{(2)}\right)^\gamma =\frac{1}{6} \left( \left(1 + \frac{1}{3^\gamma}\right) + \sqrt{12\left(1 + \frac{1}{2^\gamma}\right)^2 + \left(5 - \frac{1}{3^\gamma}\right)^2} \right).     
 \end{equation} 
which coincides with Eq.~(18) in~ \cite{Kiessling2025}.

\subsubsection{Higher Bounds $\tau_{c}^{(N)}$}

\begin{figure}
    \centering
    \includegraphics[width=1\linewidth]{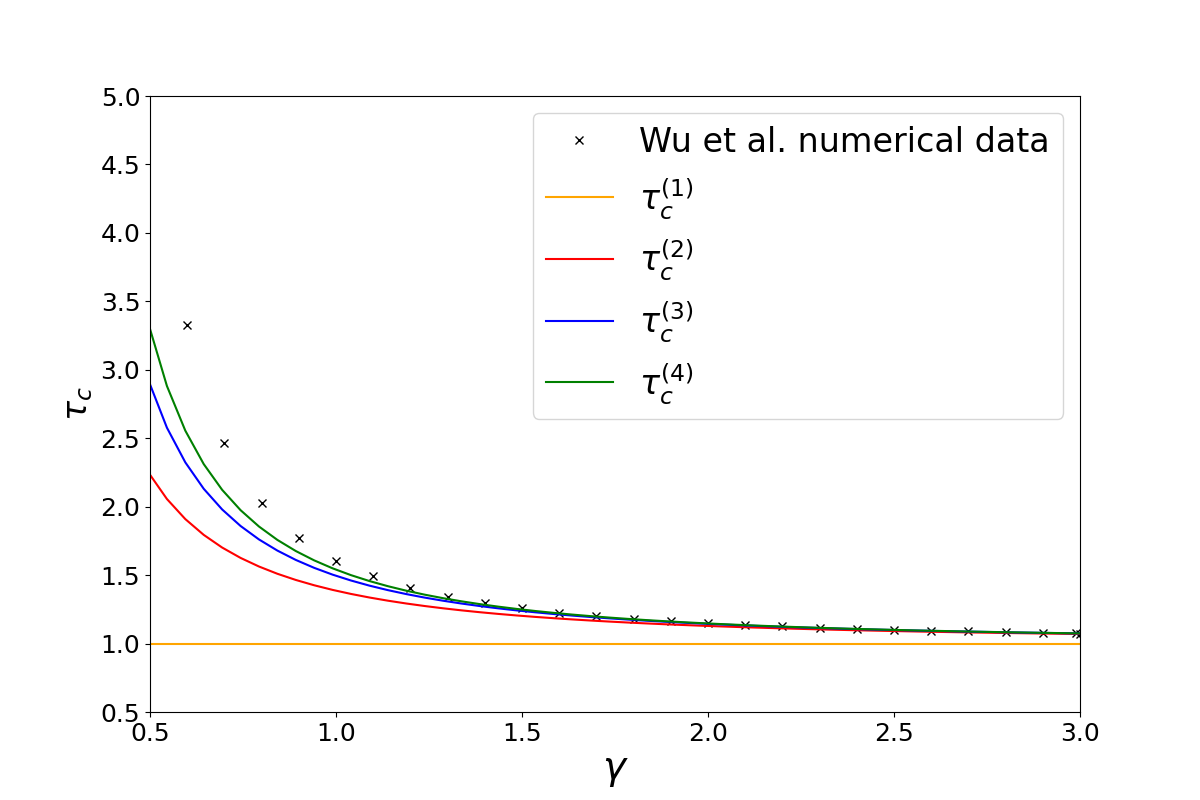}
    \caption[First Four Lower Bounds on Transition Temperature]{Comparison of the first four lowest bounds $\tau_{c}^{(1)},\tau_{c}^{(2)},\tau_{c}^{(3)},\tau_{c}^{(4)}$ versus $\gamma$, plotted alongside previous numerical solutions of the generalized Eliashberg equations from \cite{PhysRevB.99.144512} (black crosses). The graphs for the first four lowest bounds are precisely those in Fig.~(1) in \cite{Kiessling2025}, adjusted for units. The plot illustrates the convergence behavior of the bounds as $\gamma$ increases, with each successive bound providing a tighter estimate of $\tau_c$.}
    \label{fig:lowerbounds}
\end{figure}

The higher lower bounds $\tau_{c}^{(N)}$ are obtained in the same way. In principle, the $3^{rd}$ and $4^{th}$ bounds can be obtained in a closed form, which can be found explicilty in \cite{Kiessling2025}. Any $N \geq 5$, requires numerical solutions, as generally no closed form exists. The first four bounds on the transition temperature are summarized in Fig. ~\ref{fig:lowerbounds}.

\begin{figure}
    \centering
    \includegraphics[width=1\linewidth]{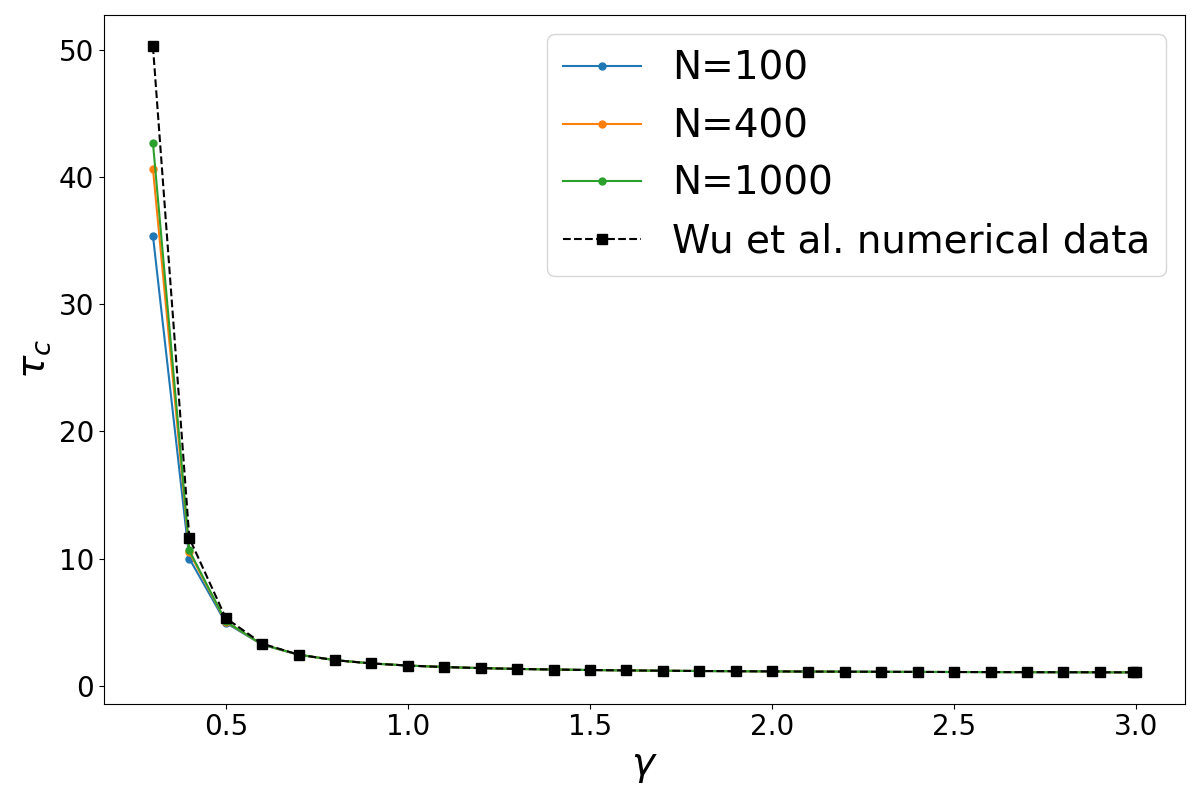}
    \caption[Convergence of the Transition Temperatures]{ The transition temperature $\tau_c$ is plotted vs $\gamma$ for different truncation values $N=100,400,1000$. for $\gamma \geq 0.5$, all truncation levels agree well with the previous numerical solutions of generalized Eliashberg equations \cite{PhysRevB.99.144512}. As $\gamma \to 0$, $\tau_c \to \infty$ as expected, which shows that the optimal truncation $N \to \infty$. All lower truncation levels serve as a "lower bound".}
    \label{fig:Nvsgamma}
\end{figure}

To investigate the effect of the truncation size, we plot the temperatures for different truncations $N=100,400,1000$ in Fig ~\ref{fig:Nvsgamma}.  We see that for $\gamma \geq 0.5$, the lowest truncation $N=100$ works reasonably well. In the region $0 < \gamma < 0.5$, we clearly see that the obtained temperature depends strongly on the truncation size. This is consistent with the condition $(N\tau)^ \gamma \ll 1$. 

\section{\label{sec:upper} Upper Bound}
In the previous section, we reproduced the lower bounds for the transition temperature. Here, we present our main result: a closed-form upper bound that significantly improves upon previous estimates.

We will use the well-known \emph{Gershgorin Circle Theorem}~\cite{Gerschgorin1931} to determine the lowest value of \( \tau \) for which all eigenvalues of the matrix are positive. The theorem is stated as follows:

\textbf{Gershgorin Circle Theorem}
Let \( H^{(N)} = (h_{ij}) \) be an \( N \times N \) matrix. For each \( i = 1, \dots, n \), define the \emph{Gershgorin disc} \( d_i \) centered at \( h_{ii} \) with radius
\[
R_i = \sum_{\substack{j=1 \\ j \neq i}}^n |h_{ij}|.
\]
Then each eigenvalue \( \lambda \) of \( H^{(N)} \) lies within at least one of the discs:
\[
\lambda \in \bigcup_{i=1}^N d_i = \bigcup_{i=1}^N \left\{ x \in \mathbb{C} : |x - h_{ii}| \leq R_i \right\}.
\]
For the detailed proof, we refer the reader to \cite{Gerschgorin1931,Varga2004,Horn2012}. We notice also that the proof holds for $N \to \infty$ under two conditions. First, the existence of a discrete eigenvalue, which we have argued in Sec.~\ref{sec:math} that it is indeed the case, and second, that the entries of the eigenvectors, i.e $\theta_n$, are bounded. Crucially, the theorem does not imply that every individual disc contains an eigenvalue. Rather, it asserts that every eigenvalue is contained within at least one disc; stated more rigorously, the complete spectrum of the matrix lies within the union of all the discs.

A key advantage of the Gerschgorin circle theorem is its broad applicability. The Gerschgorin theorem applies to any square matrix with real or complex entries, regardless of whether the matrix is symmetric or not. This makes it a highly versatile tool for estimating the location of eigenvalues in a wide variety of contexts. The theorem provides two ways to construct the disks that contain the eigenvalues: one based on rows and another on columns. For any given square matrix $H$, the \emph{row disks} are centered at the diagonal entries $h_{ii}$ with radii $R_i = \sum_{j \ne i} |h_{ij}|$, while the \emph{column disks} are centered at the same points but with radii $C_i = \sum_{j \ne i} |h_{ji}|$. For a non-symmetric matrix, these two sets of disks are generally different, but the theorem still applies to both of them.

\subsection{Simplest Case}
Consider the simplest case where we apply the theorem directly to the Hessian matrix. Since our Hessian matrix is real and symmetric, the discs become intervals on the real number line. Recall that the diagonal elements are given by Eq.~\eqref{eq:Hessiandiag}:
\[
h_{ii} = (2i+1)\tau^\gamma + 2\sum_{m=1}^{i} \frac{1}{m^\gamma} - \frac{1}{(2i+1)^\gamma},
\]
and the radii are given by the summation of off-diagonal elements and taking the limit $N \to \infty$:
\[
R_i = 2\sum_{j=1}^\infty \frac{1}{j^\gamma}- \frac{1}{(2i+1)^\gamma}.
\]

The term $ \sum_{j=1}^\infty \frac{1}{j^\gamma} = \zeta(\gamma)$ for $\gamma >1$ and diverges otherwise. It follows that each disc \( d_i \) is bounded from below by:
\begin{equation}\label{disk}
(2i+1)\tau^\gamma + 2\sum_{m=1}^{i} \frac{1}{m^\gamma} - 2\sum_{j=1}^\infty \frac{1}{j^\gamma} \leq d_i^\downarrow.
\end{equation}

For example:
\[
\begin{aligned}
d_0^\downarrow &\geq \tau^\gamma - 2\sum_{j=1}^\infty \frac{1}{j^\gamma}, \\
d_1^\downarrow &\geq 3\tau^\gamma + 2 - 2\sum_{j=1}^\infty \frac{1}{j^\gamma}, \\
d_2^\downarrow &\geq 5\tau^\gamma + 2\left(1 + \frac{1}{2^\gamma}\right) - 2\sum_{j=1}^\infty \frac{1}{j^\gamma}, \quad \text{and so on}.
\end{aligned}
\]
It is evident that for any \( \tau \geq 0 \), the lower bound of \( d_0 \) is the smallest among all \( d_i \), and thus serves as a lower bound for the union of all Gershgorin discs.

To ensure that no phase transition occurs, all eigenvalues must be positive. Let \( \lambda_N \) denote the smallest eigenvalue of \( H^{(N)} \) for sufficiently large \( N \to \infty \), and suppose \( \lambda_N \approx 0^+ \). Then, by the Gershgorin theorem:
\begin{equation}
\tau^\gamma - 2\sum_{j=1}^\infty \frac{1}{j^\gamma} \leq \lambda_n = 0^+ \quad \Rightarrow \quad \tau^\gamma < 2\sum_{j=1}^\infty \frac{1}{j^\gamma}.
\end{equation}
This implies the following upper bound for the transition temperature:
\begin{equation}\label{upperbound}
\tau_c^{\text{max}} = (2\zeta(\gamma))^\frac{1}{\gamma}, \qquad \gamma >1.
\end{equation}

This is the temperature above which all eigenvalues of any \( H \) are positive. The result in Eq.~ \eqref{upperbound} is only valid for $\gamma>1$, and does not provide good improvement at values $\gamma \leq 1.5$. We propose a new technique to improve this upper bound, while still using Gershgorin Circle Theorem. 

\subsection{Improving the Upper Bound}
Since the Gerschgorin circle theorem is not restricted to symmetric matrices, we can manipulate our original matrix to obtain another matrix with the same eigenvalues, or at least their respective signs, while tightening the Gerschgorin bounds. 

We note that as $\tau $ becomes less than $\tau_{c}$ the determinant of the Hessian changes sign. So if instead of the Hessian $H^{(N)}$ we consider another matrix
\begin{equation}\label{eq:h}
h^{(N)}=O^{-1}H^{(N)}O 
\end{equation}
where $O$ is some, say, positive definite $N\times N$ matrix,
then as $\det h^{(N)}=\det H^{(N)}$, $\tau_{c} $ --- the temperature at which the lowest eigenvalue is zero --- is the same for both $h^{(N)}$ and $H^{(N)}$.

We then can choose the matrix $O$. 

Let us take $O$ in the form $O=\text{diag}(1/p ,1/(1+p), \dots 1/(n+p),\dots 1/(N+p-1))$, where $0<p<1$ is some tunable parameter.

The matrix elements of $h^{(N)}$ are
\begin{align}
& h_{nn} = H_{nn} = (2n+1)\tau^\gamma + 2 \sum_{m=1}^{n} \frac{1}{m^\gamma} - \frac{1}{(2n+1)^\gamma} \label{eq:hdiag} \\
& h_{mn} = (m+p) H_{mn}\frac{1}{n+p}\nonumber\\
&= -(m+p)\left( \frac{1}{|m - n|^\gamma} + \frac{1}{(m + n + 1)^\gamma} \right)\frac{1}{n+p}  \label{eq:hoffdiag}
\end{align}

The Gershgorin radii are
\begin{eqnarray}
&&R^{(N)}_{m}
=
\sum_{\substack{n=0\\n\not=m}}^{N}|h_{mn}| 
\label{eq:RmN}\\
&&=
(m+p)\!\sum_{\substack{n=0\\n\not=m}}^{N}\left( \frac{1}{|m - n|^\gamma} + \frac{1}{(m + n + 1)^\gamma} \right)\frac{1}{n+p}
\nonumber
\end{eqnarray}
Then we can take the limit $N\rightarrow \infty $, as the sum converges for any $\gamma >0$. The problem becomes an optimization problem for the parameter $p$. Also, it is worth noting here that we can no longer assume that the zeroth Gershgorin disc is the lowest disc and use it to calculate the upper bound. In the appendix, we show that indeed for some values of $p$, the zeroth disc is not the lowest. However, we prove in the appendix that for $p=1/2$, the zeroth disc is the lowest disc for any $\gamma >0$. Thus, we use the zeroth disc to compute the upper bound by setting $p=1/2$.

\begin{figure}[h]
    \centering
    \includegraphics[width=1\linewidth]{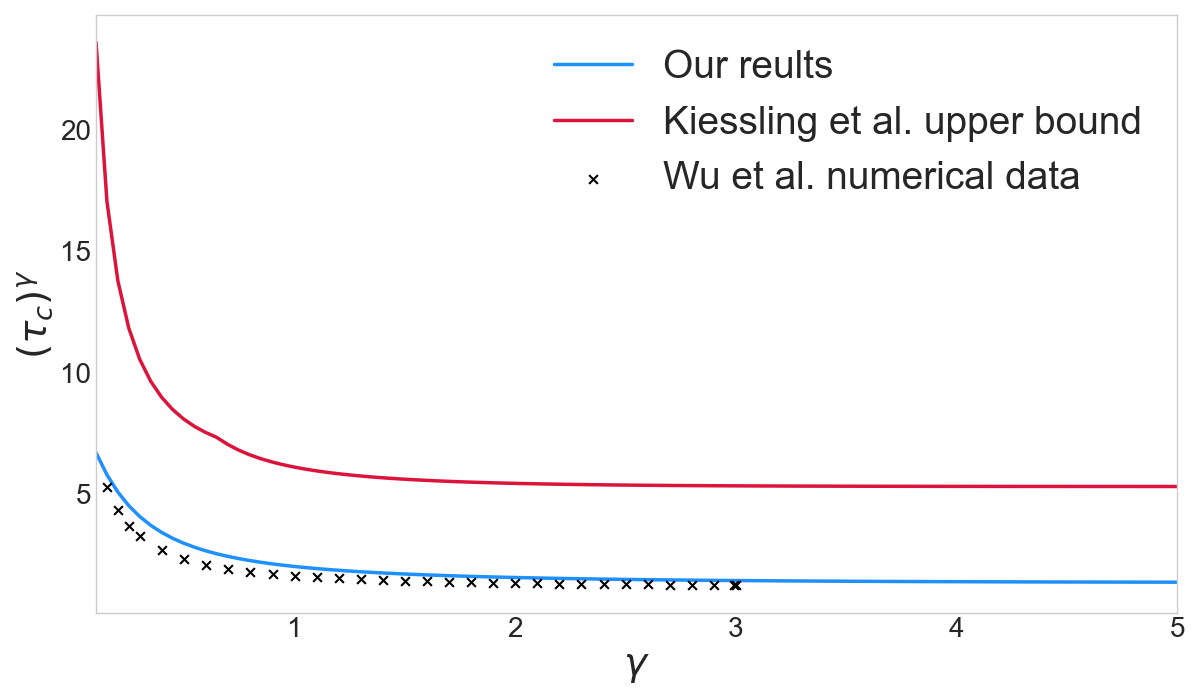}
    \caption[Comparison of Our Upper Bound vs Numerical Data]{Comparison of our results (blue line) for the upper bound of the critical temperature obtained from Eq.~\ref{eq:ourbound}   as a function of $\gamma$ in the range $0.1 \leq \gamma \leq 5$ compared to the numerical data (black crosses) from \cite{PhysRevB.99.144512} in comparison to the Upper bound obtained by Kiessling et. al \cite{Kiessling2025} (red line). Our results show a significant improvement in the upper bound. Note that we are plotting $\tau_c^\gamma$ to be able to fit both bounds in the same plot.}
    \label{fig:upper}
\end{figure}

Thus, for $p=1/2$, the expression for the upper bound is given by 
\begin{equation}\label{eq:ourbound}
\begin{split}
     \tau^{\gamma }_{\text{up}} &=\frac{1}{2}\sum_{n=1}^{\infty }\frac{1}{n^{\gamma }}\frac{2n }{(n +1/2)(n-1/2)}\\
     &= \sum_{n=0}^ \infty \left(\frac{1}{2}\right)^{2n}\zeta(\gamma+2n+1)
\end{split}
\end{equation}
which converges for any $\gamma >0$. As $\gamma \to \infty$,  $\tau_{up}^\gamma \to\frac{4}{3} =1.3333...$, close to the asymptotic behavior $\sim 1.1843$ obtained in \cite{emil2,wang}. The comparison of our results against the numerical values and Kiessling et. al's upper bound is plotted in Fig.~\ref{fig:upper}.

\section{Summary and Conclusion}\label{sec:summary}
This paper formally addresses the problem of determining the critical transition temperatures in the $\gamma$-model of quantum-critical superconductivity by performing a linear stability analysis of the Hessian matrix of the free energy functional obtained through mapping the theory into a classical spin chain. The primary challenge is that the Hessian is an infinite-dimensional operator that is not bounded, and consequently non-compact, as the digonal elements $H_{nn} \to \infty$ as $n \to \infty$. This means that a truncation to a finite matrix is not straightforward, mathematically speaking, and needs to be rigorously justified.

To establish the mathematical validity of truncating the infinite Hessian matrix, we leverage the operator-theoretic framework established in \cite{Kiessling2025}. Specifically, a congruence transformation is employed to explicitly recast the original unbounded Hessian, $H$, into a new, bounded operator, $\tilde{X}$, on the Hilbert space $\ell^2$. This transformation preserves the signs of the eigenvalues. The structure of $\tilde{X}$ as a simple term plus a compact operator, combined with Weyl's theorem, guarantees that any instability must arise from a discrete, negative eigenvalue, thereby rigorously justifying the reduction of the transition temperature search to a tractable finite $N \times N$ diagonalization problem in the limit $N \to \infty$.

With the truncation justified, we derived rigorous bounds for the first transition temperature. To establish lower bounds, we applied the Rayleigh-Ritz variational method. This guarantees that the lowest eigenvalue of any $N \times N$ truncation is greater than or equal to that of the $(N+1) \times (N+1)$ truncation, creating a monotonically increasing sequence of transition temperatures that converges from below. By solving Eq ~\ref{eq:Lower} for the transition temperatures of the $N=1, 2, 3,$ and $4$ truncations, we obtained a sequence of increasingly tight lower bounds. These results provide an independent confirmation of the findings of Kiessling et al. \cite{Kiessling2025}.

To establish upper bounds, we employed the Gerschgorin Circle Theorem \cite{Gerschgorin1931,Varga2004}, which provides a region in the complex plane guaranteed to contain all eigenvalues. An upper bound on the transition temperature is found by determining the temperature at which the lowest Gerschgorin disk touches zero. To tighten this bound, we applied an optimized similarity transformation using a diagonal matrix $O$ parameterized by a single variable, $p$. Through an analytical optimization, we found the optimal value $p=1/2$ that minimizes the resulting upper bound on the temperature. This procedure yielded a new, closed-form expression for the upper bound that represents a significant improvement over the results of Kiessling et al. \cite{Kiessling2025} and is much closer to the numerical transition temperature obtained from solving the coupled Eliashberg equations \cite{PhysRevB.99.144512}. Our results, in comparison to numerical data and current bounds in the literature are summarized in Fig.~(\ref{fig:full}) in the introduction section.

\begin{acknowledgments}
    We thank Yi-Ming Wu for sharing the numerical results from \cite{PhysRevB.99.144512}. We also thank Michael K-H Kiessling for sharing some of the numerical data from \cite{Kiessling2025}. The work of ArA was supported by the DOE-Office Of Science, DE-SC0026038.
\end{acknowledgments}

\appendix
\section{Calculating the Optimal $p$}
\begin{figure}[h!]
\centering
\includegraphics[width=\columnwidth]{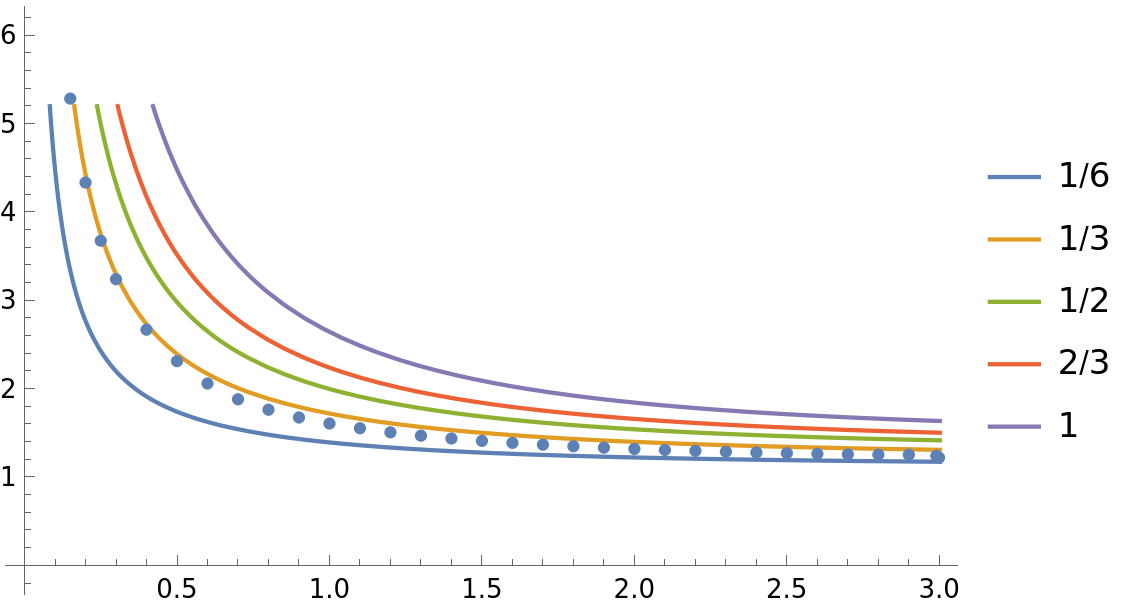}
\caption[The Bound $\tau_0$ as a Function of the Optimizing Parameter $p$.]{The upper bound  $\tau_0^ \gamma$ vs $\gamma $ for different values of $p$ shown as legends. Dots are the numerical data. It is clear that for $p=1/6$, the upper bound is below the data. It indicates that somewhere $1/2 < p < 1/3$, the assumption that the zeroth Gershgorin disc is the lowest is no longer valid.  
}
\label{fig:Diffp}
\end{figure}

Computing the lower boundaries of the Gershgorin discs $d_m ^\downarrow$, we have:
\begin{equation}
d_{m=0}^{\downarrow }
=
\tau^{\gamma }
-
p  
\sum_{n=1}^{\infty }\frac{1}{n^\gamma}
\left(
\frac{1}{n+p  }
+ 
\frac{1}{n+p-1 } 
\right) 
\label{eq:d0pAlpha}    
\end{equation}
\begin{eqnarray} 
&&
d^{\downarrow }_{m>0}
=
(2m+1)\tau^\gamma \label{eq:dmpAlpha}  \\
&&
+
\sum_{n=1}^{m }\frac{1}{n^\gamma}
\left(
2
- 
\frac{m+p }{m-n+p }
- 
\frac{m+p }{m+n+p  }
\right)
\nonumber\\
&&
-
(m+p) \!\!\!\! \sum_{n=m+1}^{\infty }\frac{1}{n^\gamma}
\left( 
\frac{1}{n+m+p  }
+ 
\frac{1}{n-m+p-1 } 
\right)
\nonumber
\end{eqnarray}
Now we introduce the temperature $\tau_{m}$ at which the $m$th Gershgorin disk touches zero.
\begin{eqnarray} 
&&
\tau_{0}^{\gamma }
=
p  
\sum_{n=1}^{\infty }\frac{1}{n^\gamma}
\left(
\frac{1}{n+p  }
+ 
\frac{1}{n+p-1 } 
\right) ,
\label{eq:tau0pAlpha} \\
&&
\tau_{m>0}^\gamma 
=
\frac{1}{2m+1}
\sum_{n=1}^{m }\frac{1}{n^\gamma}
\frac{2n^2}{(m+p)^2-n^2}
\label{eq:taumpAlpha} \\
&&
+
\frac{(m+p) }{2m+1} 
\sum_{n=m+1}^{\infty }\frac{1}{n^\gamma}
\frac{2n+p-1}{(n+m+p) (n-m+p-1) }
.
\nonumber
\end{eqnarray}
The true upper bound $\tau_{u}$ for $\tau_{c}$ for given $p$ is then
\begin{equation}\label{eq:tauBound}
\tau_{u}^{\gamma }=\max_{m}\tau^{\gamma }_{m}.
\end{equation}
Unlike the simplest case, it is not clear whether the zeroth disc $d_0$ is the lowest of all the discs for any value of $p$. In Fig.~\ref {fig:Diffp}, we plot the temperature $\tau_0^\gamma$ vs $\gamma$ for different values of $p$. We immediately see that for some values of $p$, the "upper bound $\tau_0$" is lower than the numerical data, indicating that the assumption that the zeroth disc is the lowest one is no longer true. 
This statement is also obtained from the numerical comparison of $\tau_m^\gamma$ at different $p$.  

However, for $p=1/2$, we show that $\tau_0^\gamma-\tau_{m>0}^\gamma\geq 0$ for any $m$ and is zero for $\gamma=0$. For $p=1/2$ the equations \eqref{eq:tau0pAlpha} and \eqref{eq:taumpAlpha} give 
\begin{equation}
    \tau_{0}^{\gamma } =\frac{1}{2}\sum_{n=1}^{\infty }\frac{1}{n^{\gamma }}\frac{2n }{(n +1/2)(n-1/2)}
\label{eq:kappa0:tau0}
\end{equation}
and
\begin{eqnarray} 
\tau_{m>0}^\gamma 
=
-
\frac{2}{2m+1}
\sum_{n=1}^{m }\frac{1}{n^{\gamma }} 
+
\frac{1}{2}
\sum_{n=1}^{\infty  }\frac{1}{n^{\gamma }}
\frac{1 }{n+m+1/2 } 
\nonumber \\
-
\frac{1}{2}
\sum_{n=1}^{\infty  }\frac{1}{n^{\gamma }}
\frac{1 }{n-m-1/2}
+
\sum_{n=1}^{\infty  }\frac{1}{(n+m)^{\gamma }}
\frac{1 }{n-1/2}
\label{eq:kappa0:taum} 
\end{eqnarray}
The quantity $\tau_0^\gamma -\tau_m ^\gamma$ can be written in the following form
\begin{eqnarray} 
&&\tau_{0}^{\gamma }-\tau_{m>0}^{\gamma } 
=
\frac{1}{2}
\sum_{n=1}^{m }\frac{1}{n^{\gamma }}
\frac{2}{m+1/2}
\label{eq:tau0minustaum} \\
&&
-
\frac{1}{2}\sum_{n=1}^{m }\frac{1}{n^{\gamma }} 
\frac{1 }{m+1/2-n}
+
\frac{1}{2}\sum_{n=1}^{m }\frac{1}{n^{\gamma }}
\frac{1 }{n +1/2} \nonumber \\
&&
 +
\frac{1}{2}\sum_{n=1}^{\infty }
\left(\frac{1}{n^{\gamma }}- \frac{1}{(n+m)^{\gamma }}\right)
\left( 
\frac{1 }{n-1/2}
-
\frac{1 }{n+m+1/2 }
\right)
\nonumber
\end{eqnarray}
The last line of this equality is obviously non-negative (it is zero for $\gamma =0$). Let us consider the first and the second lines. They can be written in the following form
\begin{eqnarray}
 &&\frac{1}{2}
\sum_{n=1}^{m }\frac{1}{n^{\gamma }}
\frac{2}{m+1/2}
\nonumber\\   
&&
-
\frac{1}{2}\sum_{n=1}^{m }\frac{1}{n^{\gamma }} 
\frac{1 }{m+1/2-n}
+
\frac{1}{2}\sum_{n=1}^{m }\frac{1}{n^{\gamma }}
\frac{1 }{n +1/2} \nonumber \\
&&
=
\frac{1}{2}\sum_{n=1}^{m-1 }\left(\frac{1}{n^{\gamma }}-\frac{1}{(m-n)^{\gamma }}\right)
\frac{1 }{n +1/2} \nonumber \\
&&
+
\frac{1}{m+1/2}
\sum_{n=1}^{m }\left(\frac{1}{n^{\gamma }}-\frac{1}{m^{\gamma }} \right)
\end{eqnarray}
The last term is obviously non-negative (it is zero for $m=1$ or $\gamma =0$), so let us focus on the first term. Note that for $n \in \{1, \dots, m - 1\}$ the expression in big parentheses 
is a decreasing sequence in $n$ that is reflection anti-symmetric about the 
centre of the set $\{1, \dots, m - 1\}$. This expression is multiplied by a positive, 
decreasing expression in $n$. It follows immediately that this term is positive. Alternatively, it can be written in the following form
\begin{eqnarray}
 &&
\frac{1}{2}\sum_{n=1}^{m-1 }\left(\frac{1}{n^{\gamma }}-\frac{1}{(m-n)^{\gamma }}\right)
\frac{1 }{n +1/2} \\
&&
=
\frac{1}{4}\sum_{n=1}^{m-1 }
\frac{\left((m-n)^{\gamma }-n^{\gamma } \right)\left(m-2n \right)}{n^{\gamma }(m-n)^{\gamma }(n+1/2)(m-n+1/2)}
\nonumber
\end{eqnarray}
The denominator in the expression under the sum is positive. In the numerator, there are two factors $(m-n)^{\gamma }-n^{\gamma }$ and $m-2n$. For $0<n<m$, these two factors always have the same sign: positive if $n<m/2$ and negative if $n>m/2$ (if $m$ is even and $n=m/2$, then both factors are zero). So, this sum is also non-negative (it is zero if $m=2$ or $\gamma =0$). So, taking all this together, this proves that $\tau_{0}^{\gamma }-\tau_{m>0}^{\gamma }$ is non-negative for any $m$ and is zero for $\gamma =0$.

\bibliography{apssamp}

\end{document}